\documentstyle[preprint,aps,psfig]{revtex}

\begin{document}

\draft
\preprint{
\vbox{
\hbox{TUM-T39-95-19}
\hbox{U.MD. PP\#96-054}
\hbox{DOE/ER/40762-071}
\hbox{ADP-95-57/T204}
}}

\title{Polarized deep-inelastic scattering from nuclei \protect \\
              --- a relativistic approach}

\author{G.Piller$^a$,
        W.Melnitchouk$^b$, 
        A.W.Thomas$^c$}
\address{$^a$ Physik Department,
              Technische Universit\"at M\"unchen,
              D-85747 Garching, Germany}
\address{$^b$ Department of Physics, 
              University of Maryland,
              College Park, MD 20742-4111}
\address{$^c$ Department of Physics and Mathematical Physics and \\
              Institute for Theoretical Physics,
              University of Adelaide, Adelaide 5005,
              Australia}
\maketitle

\begin{abstract}
We discuss spin-dependent, deep-inelastic scattering from 
nuclei within a covariant framework.
In the relativistic impulse approximation this is described 
in terms of the amplitude for forward, virtual-photon scattering from an
off-mass-shell nucleon.
The general structure of the off-shell nucleon hadronic tensor
is derived, and the leading behavior of the off-shell nucleon
structure functions computed in the Bjorken limit.
The formalism, which is valid for nucleons bound inside 
nuclei with spin 1/2 or 1, is applied to the 
case of the deuteron.
\end{abstract}

\pacs{PACS numbers: 13.60.Hb, 24.10.Jv, 25.30.Fj   \\ \\ \\
      To appear in {\em Phys.Rev.C.}}

\section{Introduction}

The role of special relativity in nuclei has been an important 
consideration in recent years in the pursuit of consistent 
descriptions of nuclear electromagnetic processes at high 
energies \cite{RLTV}. 
Evidence suggests, for instance, that non-relativistic models 
are inadequate  
to account for elastic form factors of few-nucleon systems at 
large momentum transfers, or to provide a quantitative description 
of proton--nucleus cross sections over the full energy range.
While meson degrees of freedom almost certainly  
play some role in these discrepancies, for a full and consistent
theoretical description that does not manifestly break Lorentz 
covariance, one must also incorporate effects that arise from 
the off-mass-shell nature of the bound nucleons.

A process in which effects associated with the off-mass-shell 
deformation of the nucleon structure function may also have an 
impact is deep-inelastic scattering (DIS)
of leptons from nuclei.
Until recently, the issue of nucleon off-shellness has 
largely been ignored in this problem, even in 
calculations based on the 
so-called relativistic impulse approximation.
Often the only relativistic corrections made are kinematic, 
without consideration of the dynamics that may be affected 
when the nucleon in a nuclear medium is off-shell.

The in-medium modification of the nucleon structure in unpolarized
scattering was discussed in Refs.\cite{MST,KPW}, and, as the only example
for which relativistic nuclear wave functions have been calculated, 
a quantitative study of the effects for a deuterium nucleus made 
in Ref.\cite{MSTD}.
The deuterium analysis was extended to polarized processes in
Ref.\cite{MPT}, where it was found that, while small for moderate 
values of Bjorken-$x$, the off-shell effects can become sizeable 
in the region of relativistic kinematics, $x \agt 0.8$
(see also Ref.\cite{KMPW} for a non-relativistic treatment).
Apart from the appreciation of the importance of relativity in nuclei 
per se, there is also a practical need to understand the 
role of off-shell effects in light nuclei such as deuterium or 
helium.
In the absence of free neutron targets, these nuclei are the only 
sources of information on the spin structure of the neutron in DIS,
which is essential for testing fundamental QCD sum rules, such as
the Bjorken sum rule.

In this paper we present a full derivation of the modification of 
the nucleon structure in the off-mass-shell region for spin-dependent
DIS.
A brief outline for the special case of massless quarks was given 
previously in Ref.\cite{MPT}. 
Here we extend the discussion in \cite{MPT} to the most general case, 
including mass terms.
In Section II we analyze the most general form of the antisymmetric 
part of the off-shell hadronic tensor which enters the calculation 
of the spin-dependent cross section.  
Any relativistic model of polarized DIS from nuclei within the 
impulse approximation must be consistent with the symmetry 
constraints on the truncated hadronic tensor derived here.
Furthermore, we calculate the scaling behavior of the expansion 
coefficients of the off-shell tensor in leading twist.
This enables us to present for the first time a model-independent proof
of gauge invariance of the off-shell, spin dependent hadronic tensor.

Working within the nuclear impulse approximation (i.e. neglecting
effects due to final state interactions), in Section III we present 
a model independent result for the nuclear structure function 
$g_1^A$ in terms of the off-shell nucleon tensor and the off-shell 
nucleon--nucleus scattering amplitude.
The impulse approximation assumes that inelastic scattering from 
nuclei proceeds via incoherent scattering from individual nucleons, 
which is believed to be a good approximation if one is sufficiently 
far away from the small-$x$ region.
The formal results are valid for spin 1/2 and spin 1 nuclei.
(They can also be applied to the case of DIS from a nucleon 
dressed by a meson cloud \cite{SST,MTV,KUM,JUEL}.) 
Furthermore, we investigate the conditions for the validity of the 
usual convolution model \cite{WOL,FS,CIOFI,SAUER,TOK,KAP}, which 
involves factorization of subprocesses at the cross section 
(rather than at the amplitude) level.

Using these results, in Section IV we illustrate the application 
of the formalism to the special case of the deuteron.
Some of the numerical results, obtained in the massless quark limit,
have been presented in Ref.\cite{MPT}. 
For completeness, we compare here the results for the proton and 
deuteron structure functions with the latest available data on 
$g_1^p$ and $g_1^D$, as well as with unpolarized quark distributions.
Finally, in Section VI we make some concluding remarks.

\section{Polarized Nucleon Structure Functions}

Here we analyze the general structure of the amplitude for the
forward scattering of a virtual photon from a polarized, off-mass-shell
nucleon. (In fact, the formalism is valid for any spin-1/2 fermion
with substructure.)
Recall that the antisymmetric part of the hadronic tensor for an
on-shell nucleon in terms of the structure functions $g_1^N$ and
$g_2^N$ is written as:
\begin{equation}
\label{WmunuNdef}
M \,W_{\mu\nu}^N(p,s,q)
= i { M \over p\cdot q }
\epsilon_{\mu \nu \alpha \beta}\ q^{\alpha}
\left( s^{\beta}\ g_1^N(p,q)\
    +\ \left( s^{\beta} - { s\cdot q \over p\cdot q } p^{\beta}
       \right) g_2^N(p,q)
\right),
\end{equation}
where $p$ and $q$ are the four-momenta of the nucleon and photon,
respectively. 
Here $M$ stands for the nucleon mass, and $s$ is the 
nucleon polarization vector.
Since we will be interested in the leading twist components
of the structure functions only, we will not discuss the
structure function $g_2$, which contains both twist-2 and twist-3
contributions.

Our aim will be to generalize the tensor structure in
$W_{\mu\nu}^N$ to describe deep-inelastic scattering from 
an off-shell nucleon,
namely one with $p^2 \not= M^2$.
We start with the observation that the antisymmetric nucleon 
tensor can be written as:
\begin{eqnarray}
\label{WmunuN}
 M\ W_{\mu\nu}^N(p,s,q) = \bar u(p,s) \, 
  \widehat G_{\mu\nu}(p,q) \, u(p,s),
\end{eqnarray}
where $\widehat G_{\mu\nu}(p,q)$ is the ``truncated'' nucleon tensor,
whose Dirac structure represents deep-inelastic scattering from a 
generally off-shell nucleon, and $u(p,s)$ is the free Dirac spinor for 
a nucleon with momentum $p$ and spin $s$. 
In the following we give the general expression for the 
truncated tensor $\widehat G_{\mu\nu}$, which satisfies 
the discrete symmetries, in terms of a number of 
``truncated coefficient functions'', or off-shell nucleon 
structure functions.
Following this, we derive expressions for the off-shell
structure functions in the Bjorken limit.

\subsection{Truncated Nucleon Structure Functions}

In analyzing the off-shell nucleon structure, it will be convenient
to expand the truncated nucleon tensor $\widehat G_{\mu\nu}$ in terms
of independent basis tensors, such that $\widehat G_{\mu\nu}$ is 
invariant under parity and time reversal.
These constraints can be summarized by the following conditions: 
\begin{mathletters}
\label{Ghatsym}
\begin{eqnarray}
\widehat G_{\mu\nu}(p,q)
&=& {\cal P}\
    \widehat G^{\mu\nu}(\widetilde p,\widetilde q)\
    {\cal P}^{\dagger}, \\
\widehat G_{\mu\nu}(p,q)
&=& \left( {\cal T}\
           \widehat G^{\mu\nu}(\widetilde p,\widetilde q)\
           {\cal T}^{\dagger}
    \right)^*,
\end{eqnarray}
where ${\cal P}$ and ${\cal T}$ are the parity and 
time reversal operators, respectively. 
(In the Dirac representation they are expressed in terms of the 
Dirac matrices as:
${\cal P} = \gamma_0$ and
${\cal T} = -i \gamma_5 {\cal C}$,
where ${\cal C} = i \gamma^2 \gamma_0$ is the charge
conjugation operator.)
We also use the notation
$\widetilde p_{\mu} = p^{\mu}$ and
$\widetilde q_{\mu} = q^{\mu}$ to distinguish covariant and 
contravariant four-vectors.
The truncated tensor must also be hermitian, which requires that
(note that $\widehat G_{\mu\nu} = - \widehat G_{\nu\mu}$):
\begin{eqnarray}
\widehat G_{\mu\nu}(p,q)
&=& \gamma_0 \widehat G_{\nu\mu}^{\dagger}(p,q) \gamma_0.
\end{eqnarray}
\end{mathletters}%

According to the constraints in Eqs.(\ref{Ghatsym}) the most general
form for the truncated tensor must be \cite{MPT,KMPW}:
\begin{eqnarray}
\label{Ghatex}
\widehat{G}_{\mu\nu}(p,q)
&=& i \epsilon_{\mu \nu \alpha \beta}\ q^{\alpha}
\left( p^{\beta} \not\!p \gamma_5\, G_{(p)}\
    +\ p^{\beta} \not\!q \gamma_5\, G_{(q)}
    +\ \gamma^{\beta} \gamma_5\, G_{(\gamma)}\
\right.                                         \nonumber\\
& & \hspace*{0cm} +\
\left. i \sigma^{\beta\lambda} p_{\lambda} \gamma_5 G_{(\sigma p)}\
    +\ i \sigma^{\beta\lambda} q_{\lambda} \gamma_5 G_{(\sigma q)}\
    +\ i p^{\beta} \sigma^{\lambda\rho}\ p_{\lambda} q_{\rho}
         \gamma_5 G_{(\sigma pq)}
\right), 
\end{eqnarray}
where the coefficients $G_{(i)}$ are scalar functions
of $p$ and $q$.
In Eq.(\ref{Ghatex}) we have listed only those structures which
lead to a gauge-invariant truncated tensor,
i.e. $q^{\mu} \widehat G_{\mu \nu} 
= q^{\nu} \widehat G_{\mu \nu} = 0$.
In Sec. II B we will show that, at least in the Bjorken limit,
terms such as 
$\epsilon_{\mu \nu \alpha \beta} p^{\alpha} \gamma^{\beta}$
that do not satisfy this condition, are absent.

The structure function $g_1^N$ of an on-shell nucleon
is obtained by multiplying $\widehat G_{\mu\nu}$ with the nucleon
Dirac spinors (see Eq.(\ref{WmunuN})) and with the  projection 
operator:  
\begin{equation}
P_{\mu\nu} = i \epsilon_{\mu\nu\alpha\beta} \,q^{\alpha}\,
          \frac{p\cdot q}{s\cdot q} \,
          \frac{s\cdot q M^2 s^{\beta} - p\cdot q p^{\beta} }
          { 2 M \left[ M^2 (q^2 + (s\cdot q)^2) - (p\cdot q)^2\right]}.
\end{equation}
As a result we get $g_1^N(x)$ as the on-shell limit 
($p^2 \rightarrow M^2, y \rightarrow 1$) of the off-shell 
structure function
$\widetilde g_1^N(x/y,p^2)$, defined as:
\begin{eqnarray}
\label{g1N}
\widetilde g_1^N \left( {x \over y}, p^2 \right)
&=& 2 p \cdot q\
\left( p \cdot q\ G_{(q)} + G_{(\gamma)}
     + M\ G_{(\sigma p)} - M p \cdot q\ G_{(\sigma pq)}
\right),
\end{eqnarray}
where $x/y = Q^2/2 p\cdot q$, and the $Q^2$ dependence in 
$g_1^N$ has been suppressed.
The definition of the generalized structure function, 
$\widetilde g_1^N$ in Eq.(\ref{g1N}), will turn out to be 
useful when we discuss the nuclear spin-structure function, 
$g_{1}^A$, in Section III.

\subsection{Truncated Functions
            in the Bjorken limit} 

In this section we discuss the scaling properties of the coefficient
functions $G_{(i)}$, and the question of the gauge-transformation
properties of the truncated tensor.
We work throughout in the Bjorken limit 
($Q^2, p \cdot q \rightarrow \infty$, $Q^2/p\cdot q$ fixed), 
and consider only the
leading twist contributions to the hadronic tensor.
In this limit, the tensor $\widehat{G}_{\mu\nu}$ can be written as: 
\begin{eqnarray}
\label{Ghatlead}
\widehat{G}_{\mu\nu}(p,q)
&=& \int d\widetilde k\
    {\rm Tr}
    \left[ {\cal H}(p,k)\ r_{\mu\nu}(k,q) \right],
\end{eqnarray}
where the trace is over quark indices.
In Eq.(\ref{Ghatlead}) the antisymmetric tensor $r_{\mu\nu}$
describes the hard, photon--quark interaction, and is given by:
\begin{mathletters}
\begin{eqnarray}
r_{\mu\nu}(k,q)
&=& (\not\!k\ +\ m)
    \gamma_{\mu} (\not\!k\ + \not\!q\ + m) \gamma_{\nu}
    (\not\!k\ +\ m)\
 -\ (\mu \leftrightarrow \nu),                   \\
&\equiv& A_{\mu\nu\alpha}(k,q)\ \gamma_5 \gamma^{\alpha}\
      +\ A_{\mu\nu\alpha\beta}(k,q)\ \sigma^{\alpha\beta},
\end{eqnarray}
where
\begin{eqnarray}
A_{\mu\nu\alpha}(k,q)
&=& i \left( q^2 \epsilon_{\mu\nu k\alpha}\
          +\ (k^2-m^2)\ \epsilon_{\mu\nu q\alpha}\
          -\ 2\ (k_{\mu} \epsilon_{kq\alpha\nu}
                -k_{\nu} \epsilon_{kq\alpha\mu})
      \right),                                   \\
A_{\mu\nu\alpha\beta}(k,q)
&=& -\ i m \left( q^2\ g_{\mu\alpha}\ g_{\nu\beta}\
            +\ 2 q_{\alpha}\ (k_{\mu} g_{\nu\beta}
                             -k_{\nu} g_{\mu\beta})
        \right).
\end{eqnarray}
\end{mathletters}%
Here $k$ is the interacting quark four-momentum, and
$m$ is its mass. We use the notation
$\epsilon_{\mu\nu k q } \equiv \epsilon_{\mu\nu\alpha\beta}
k^{\alpha} q^{\beta}$.
(The complete forward scattering amplitude would also contain 
a crossed photon process which we do not consider here,  
since in the subsequent model calculations  
we focus on valence quark distributions.)
The function ${\cal H}(k,p)$ represents the soft quark--nucleon
interaction.
Since one is calculating the imaginary part of the forward
scattering amplitude, the integration over the quark momentum $k$
is constrained by $\delta$-functions which put both 
the scattered quark and the  non-interacting 
spectator system on-mass-shell:
\begin{eqnarray}
d\widetilde k
&\equiv& { d^4k \over (2\pi)^4 }
         { 2\pi \delta((k+q)^2 - m^2)\
           2\pi \delta((p-k)^2 - m_S^2) \over
           (k^2-m^2)^2 },
\end{eqnarray}
where $m_S^2 = (p-k)^2$ is the invariant mass squared of 
the spectator system.

Taking the trace over the quark spin indices we find:
\begin{eqnarray}
{\rm Tr} \left[ {\cal H}\ r_{\mu\nu} \right]
&=& A_{\mu\nu\alpha} H^{\alpha}\
 +\ A_{\mu\nu\alpha\beta} H^{\alpha\beta},
\end{eqnarray}
where $H_{\alpha}$ and $H_{\alpha\beta}$ are vector
and tensor coefficients, respectively.
The general structure of $H_{\alpha}$ and $H_{\alpha\beta}$
can be deduced from the transformation properties of the truncated
nucleon tensor $\widehat G_{\mu\nu}$ and the tensors
$A_{\mu\nu\alpha}$ and $A_{\mu\nu\alpha\beta}$.
Namely, from
$A_{\mu\nu\alpha}^*(k,q)
=A_{\nu\mu\alpha}(k,q)$ and
$A^{\mu\nu\alpha}(\widetilde k,\widetilde q)
=- A_{\mu\nu\alpha}(k,q)$,
we have:
\begin{mathletters}
\label{Hsym}
\begin{eqnarray}
H^{\alpha}(p,k)
&=& - {\cal P} H_{\alpha}(\widetilde p,\widetilde k)
      {\cal P}^{\dagger}, \\
H^{\alpha}(p,k)
&=& \left( {\cal T} H_{\alpha}(\widetilde p,\widetilde k)
           {\cal T}^{\dagger}
    \right)^*,                                           \\
H^{\alpha}(p,k)
&=& \gamma_0 H^{\alpha \dagger}(p,k) \gamma_0.
\end{eqnarray}
\end{mathletters}%
Similarly, since
$A_{\mu\nu\alpha\beta}^*(k,q)
=A_{\nu\mu\alpha\beta}(k,q)$ and
$A^{\mu\nu\alpha\beta}(\widetilde k,\widetilde q)
=A_{\mu\nu\alpha\beta}(k,q)$,
one finds:
\begin{mathletters}
\label{HHsym}
\begin{eqnarray}
H^{\alpha\beta}(p,k)
&=& {\cal P} H_{\alpha\beta}(\widetilde p,\widetilde k)
    {\cal P}^{\dagger}, \\
H^{\alpha\beta}(p,k)
&=& - \left( {\cal T} H_{\alpha\beta}(\widetilde p,\widetilde k)
             {\cal T}^{\dagger}
      \right)^*,                                           \\
H^{\alpha\beta}(p,k)
&=& \gamma_0 H^{\alpha\beta \dagger}(p,k) \gamma_0.
\end{eqnarray}
\end{mathletters}%
With these constraints, the tensors $H_{\alpha}$ and
$H_{\alpha\beta}$ can be projected onto Dirac and Lorentz
bases as follows:
\begin{mathletters}
\label{Hdef}
\begin{eqnarray}
H_{\alpha}
&=& p_{\alpha} \gamma_5 (\not\!p\ g_1\ +\ \not\!k\ g_2)\
 +\ k_{\alpha} \gamma_5 (\not\!p\ g_3\ +\ \not\!k\ g_4)\
 +\ i \gamma_5 \sigma_{\lambda\rho}\ p^{\lambda} k^{\rho}
    (p_{\alpha}\ g_5\ +\ k_{\alpha}\ g_6)\              \nonumber\\
&+& \gamma_{\alpha} \gamma_5\ g_7\
 +\ i \gamma_5 \sigma_{\lambda\alpha}
    (p^{\lambda}\ g_8\ +\ k^{\lambda}\ g_9),                    \\
H_{\alpha\beta}
&=& (p_{\alpha} k_{\beta} - p_{\beta} k_{\alpha})\
    \sigma_{\lambda\rho} p^{\lambda} k^{\rho}\ f_1\
 +\ (p_{\alpha} \sigma_{\lambda\beta} - p_{\beta} \sigma_{\lambda\alpha})
    (p^{\lambda}\ f_2\ +\ k^{\lambda}\ f_3)\            \nonumber\\
&+& (k_{\alpha} \sigma_{\lambda\beta} - k_{\beta} \sigma_{\lambda\alpha})
    (p^{\lambda}\ f_4\ +\ k^{\lambda}\ f_5)\
 +\ \sigma_{\alpha\beta} f_6\
 +\ \epsilon_{\lambda\rho\alpha\beta} p^{\lambda} k^{\rho} \gamma_5\
    (\not\!p\ f_7\ +\ \not\!k\ f_8)\                    \nonumber\\
&+& \epsilon_{\lambda\rho\alpha\beta}\ \gamma_5 \gamma^{\rho}\
    (p^{\lambda}\ f_9\ +\ k^{\lambda}\ f_{10}),
\end{eqnarray}
\end{mathletters}%
where the functions $g_{1\cdots9}$ and $f_{1\cdots10}$ are scalar
functions of $p$ and $k$.

Performing the integration over $k$ in Eq.(\ref{Ghatlead}) and using
Eqs.(\ref{Hdef}), we obtain expressions for the truncated structure
functions $G_{(i)}$ in terms of the non-perturbative coefficient 
functions $f_{i}$ and $g_{i}$. 
The explicit forms of these are given in Appendix \ref{Gi}.
{}From Eq.(\ref{Ghatex}) we then obtain the 
leading twist contributions  to the truncated nucleon tensor 
$\widehat G_{\mu \nu}$.  
It is important to note that 
at leading twist the non-gauge invariant contributions to 
$\widehat G_{\mu\nu}$ vanish, so that the expansion in
Eq.(\ref{Ghatex}) is the most general one which is consistent
with the gauge invariance of the hadronic tensor.

\section{Nuclear Structure Functions}

Our discussion of polarized deep-inelastic scattering from
nuclei is restricted to the nuclear impulse approximation, illustrated
in Fig.1.
Nuclear effects which go beyond the impulse approximation include
final state interactions between the nuclear debris of the struck
nucleon \cite{SAIT}, corrections due to meson exchange currents
\cite{KUL,KAP3,MT} and nuclear shadowing (see \cite{PRW,MTSha}
and references therein).
Since we are interested in the medium and large-$x$ regions, 
coherent multiple scattering effects, which lead to nuclear 
shadowing for $x \alt 0.1$, will not be relevant.   
In addition, it has been argued \cite{KMPW} that meson exchange 
currents are less important
in polarized deep-inelastic scattering than in the unpolarized
case since their main contribution comes from pions.

Within the impulse approximation, deep-inelastic
scattering from a polarized nucleus with spin 1/2 or 1
is then described as a two-step process, in terms of the
virtual photon--nucleon interaction,
parametrized by the truncated antisymmetric nucleon tensor
$\widehat G_{\mu\nu}(p,q)$,
and the polarized nucleon--nucleus scattering amplitude
$\widehat A(p,P,S)$.  
The antisymmetric part of the nuclear
hadronic tensor can then be written as:
\begin{eqnarray}
M_A \,W_{\mu\nu}^A(P,S,q)
&\equiv& i { M_A \over P\cdot q }
\epsilon_{\mu \nu \alpha \beta}\ q^{\alpha}
\left( S^{\beta}\ g^A_{1}(P,q)\
    +\ \left( S^{\beta} - { S\cdot q \over P\cdot q } P^{\beta}
       \right) g^A_{2}(P,q)
\right),                                         \\
&=& 
    \int \frac{ d^4 p }{ (2\pi)^4 }
    {\rm Tr} \left[ \widehat{A}(p,P,S)\ \
                    \widehat{G}_{\mu\nu}(p,q)
             \right].   \label{Wmunu}
\end{eqnarray}
Here $P$, $p$ and $q$ are the nucleus, off-shell nucleon
and photon momenta, respectively.
For a spin 1/2 nucleus, such as $^3He$, the vector
$S^{\alpha}$ ($S^2=-1, P\cdot S=0$) is the nuclear spin vector,
while for spin 1 targets, such as deuterium,
$S^{\alpha}$ is defined in terms of polarization vectors
$\varepsilon_{\alpha}^m$ \cite{JHM}:
\begin{equation}
S^{\alpha}(m)
\equiv  -i\ \epsilon^{\alpha\beta\lambda\rho}\
            \varepsilon^{m *}_{\beta}\ \varepsilon^{m}_{\lambda}\
            P_{\rho} / M_A,
\label{eq:Seps}
\end{equation}
where $m=0,\pm 1$ is the spin projection. 
Using the fact that the nuclear recoil states are on-shell,
one finds for the nucleon--nucleus amplitude:
\begin{equation}
\widehat A(p,P,S) = \sum_R 
\left \langle P-p,R\right| N(0)\left | P,S \right\rangle 
\left\langle P,S\right| \overline N(0)\left | P-p,R\right\rangle
2 \pi \,\delta \left ( (P-p)^2 - M_R^2\right).
\label{eq:A2}
\end{equation}
The sum is taken over all possible nuclear recoil states $R$
with momentum $P-p$ and invariant mass $M_R$,  
and $N(0)$ stands for the nucleon field operator at the origin. 
The amplitude $\widehat A(p,P,S)$ can furthermore be expanded in 
the nucleon Dirac space as:         
\begin{equation}
\label{ApPS}
\widehat A(p,P,S) = \sum_R 2\pi\,\delta\left((P-p)^2 - M_R^2\right)
 \left(
        \gamma_5\gamma_{\alpha} {\cal A}_R^{\alpha} +
        \sigma_{\alpha\beta} {\cal B}_R^{\alpha \beta}\right),
\end{equation}
where the pseudovector and tensor structure 
${\cal A}_R^{\alpha}$ and ${\cal B}_R^{\alpha \beta}$
are functions of $p,P$ and $S$. 
Other structures do not contribute to $W^A_{\mu\nu}$ and are 
omitted. 
For example, a pseudoscalar term 
$(\sim \gamma_5)$ is forbidden by hermiticity and time 
reversal invariance of $W^A_{\mu\nu}$. 
{}From Eqs.(\ref{Ghatex}), (\ref{Wmunu}) and (\ref{ApPS}) we obtain,
after the appropriate projection:
\begin{eqnarray}
g^A_{1}(x)
&=& \frac{P\cdot q}{4 \pi^2 M_A S\cdot q}
    \sum_R \int dy\,dp^2\,
    \left[ q\cdot {\cal A}_R\, 
    \left( p\cdot q \,G_{(q)} + G_{(\gamma)} \right)\,
     + p\cdot {\cal A}_R  \,p\cdot q \,G_{(p)}
    \right.                                             \nonumber \\
& & \hspace{5cm}
    \left.+ {\cal T \cdot B}_R\,M\,  
    \left( G_{(\sigma p)} - p\cdot q \,G_{(\sigma p q)}
    \right)
    \right], 
\label{eq:g1A}
\end{eqnarray}
with the tensor
\begin{eqnarray}
{\cal T}_{\alpha\beta}
&=& \frac{1}{(S\cdot q)^2 M_A^2 - (P\cdot q)^2}
    \left( p\cdot q \,P\cdot q \,\epsilon_{\alpha \beta P q}
         + M_A^2 \,p\cdot q \,S\cdot q \,\epsilon_{\alpha \beta q S}
    \right.                                             \nonumber \\
& & \left.
         + P\cdot q\,q_{\alpha} \,\epsilon_{\beta p P q}
         - P\cdot q\,q_{\beta} \,\epsilon_{\alpha p P q}
         + M_A^2 \,S\cdot q\,q_{\alpha} \,\epsilon_{\beta p q S}
         - M_A^2 \,S\cdot q\,q_{\beta} \,\epsilon_{\alpha p q S}
     \right).
\end{eqnarray}
Here $x=Q^2/2 P\cdot q$ is the Bjorken scaling variable and
$y = p\cdot q/P\cdot q$ is the light-cone fraction of the deuteron
momentum carried by the struck nucleon.
Equation (\ref{eq:g1A}) shows that factorization of the nuclear 
structure function $g_1^A$ into nuclear 
$\left({\cal A}_R^{\alpha},\,{\cal B}_R^{\alpha \beta} \right)$
and nucleon $\left( G_{(i)} \right )$ parts, as would be required 
for convolution, is not possible. 
In addition, the presence of the structure $G_{(p)}$ in $g_1^A$, 
but not in $g_1^N$ (see Eq.(\ref{g1N})), leads to explicit 
convolution breaking. 
Convolution can be recovered, however, in the non-relativistic 
limit \cite{KMPW}, as will be shown explicitly in the next section 
for the case of the deuteron.

\section{Deuteron}

The general results of the previous section 
can now be applied to the case of a deuteron target. 
Consistent with the standard treatments of the relativistic 
deuteron bound state problem, the deuteron recoil state is 
taken to be a single nucleon. 
{}From Eq.(\ref{eq:A2}) one therefore obtains the nucleon--deuteron 
scattering amplitude:   
\begin{equation}
\widehat A(p,P,S) = \sum_s 
\left \langle P-p,s\right| N(0)\left | P,S \right\rangle 
\left\langle P,S\right| \overline N(0)\left | P-p,s \right\rangle
2 \pi \,\delta \left ( (P-p)^2 - M^2\right),
\label{eq:AD1}
\end{equation}
where the sum is taken over the spin $s$ of the recoil nucleon. 
The deuteron--nucleon matrix elements in Eq.(\ref{eq:AD1}) are related 
to the deuteron--nucleon vertex function $\Gamma_D$ via \cite{BG}
(see also Ref.\cite{KT}):
\begin{equation}
\left\langle P-p,s\right| N(0)\left | P,S\right\rangle 
 = \frac{1}{\not\!p - M}\,\Gamma_D^{\alpha}(P,p)\, 
\varepsilon_{\alpha}^m 
\,{\cal C}\,\overline u^T(P-p,s).
\label{eq:GammaD}
\end{equation}
The deuteron polarization vector, $\varepsilon_{\alpha}^m$, is 
related to the spin vector $S$ by  Eq.(\ref{eq:Seps}). 
${\cal C}$ is the charge conjugation operator, 
and $u(P-p,s)$ is the Dirac 
spinor of the recoil nucleon with momentum $P-p$ and spin $s$. 
In terms of the deuteron--nucleon vertex function,   
Eqs.(\ref{eq:AD1},\ref{eq:GammaD}) give:
\begin{eqnarray}
\widehat A(p,P,S) &=&  \nonumber \\
&&\hspace{-2.cm} \varepsilon_{\alpha}^m \epsilon_{\beta}^{m\,*} 
(\not\!p - M)^{-1} \Gamma_D^{\alpha}(P,p) 
    (\not\!P - \not\!p - M)\
    \overline{\Gamma}_D^{\beta}(P,p)\
    (\not\!p - M)^{-1}\  
    2 \pi \delta \left ( (P-p)^2 - M^2\right),   
\end{eqnarray}
with $\overline {\Gamma}_D = \gamma_0 \Gamma_D^{\dagger}\gamma_0$.

To be specific, we will use in our numerical calculations  
the relativistic
vertex function from Ref.\cite{BG}. 
In this work $\Gamma_D$ is represented through a relativistic 
deuteron wave function which contains $S$- and $D$-state 
components $u$ and $w$, and also triplet and singlet 
$P$-state contributions $v_t$ and $v_s$. 
After choosing the polarization of the deuteron one 
can express the pseudovector and tensor components,  
${\cal A}^{\alpha}$ and ${\cal B}^{\alpha \beta}$, 
of the deuteron-nucleon amplitude (\ref{ApPS}) in terms of the 
deuteron-nucleon vertex function --- or equivalently through the 
deuteron wave function components 
(see Eq.(\ref{Aonq}) below). 
Then Eq.(\ref{eq:g1A}) yields the deuteron spin structure 
function $g_1^D$.

We will follow this procedure in the deuteron rest frame 
where the photon momentum is chosen along the $-\hat z$ direction,   
$q = \left(q_0; {\bf 0}_T,-|{\bf q}|\right)$. 
Furthermore, we are free to fix the deuteron spin projection 
to be $m=+1$. 
As in the general case in Eq.(\ref{eq:g1A}), one does not 
obtain a result compatible with exact convolution. 
However, by writing $q\cdot {\cal A}$, $p\cdot {\cal A}$ and
${\cal T\cdot  B}$ in terms of the relativistic deuteron wave 
function, we find that all non-factorizable corrections to 
convolution are at least of order $(v/c)^2$, or involve 
relativistic $P$-state wave functions. 
This is easily seen by separating $q\cdot {\cal A}$ and 
${\cal T \cdot B}$ into an ``on-shell'' part, which is 
proportional to the non-relativistic deuteron wave function 
(see Eq.(\ref{PsiNR}) below), and an off-shell component:
$q\cdot {\cal A}\equiv q\cdot {\cal A}_{ON} + q\cdot {\cal A}_{OFF}$, and 
${\cal T\cdot B} \equiv {\cal T\cdot B}_{ON} + {\cal T\cdot B}_{OFF}$,
where
\begin{eqnarray} \label{Aonq}
q\cdot {\cal A}_{ON} 
&=& {\cal T\cdot B}_{ON} \nonumber \\
&=& 2 \pi^2 P \cdot q M
    \left[ u^2 + \left( 1 - 3 \cos^2\theta \right) {u w\over\sqrt{2}}
               - \left( 1 - {3\over 2} \cos^2\theta \right) w^2\
        +\ {p_z \over M} \left( u - { w \over \sqrt{2} } \right)^2
    \right],
\end{eqnarray}
with 
$p_z = |{\bf p}| \cos\theta$,
$E_p = \sqrt{ M^2 + {\bf p}^2 }$ and
$\cos\theta = (y M_D - p_0) / |{\bf p}|$.
The ``off-shell'' components $q\cdot {\cal A}_{OFF}$ and 
${\cal T\cdot B}_{OFF}$, and also $p\cdot {\cal A}$ are given in 
Appendix \ref{OFF}. They are either of higher order in $(v/c)$ 
compared with the 
leading ``on-shell'' contribution in Eq.(\ref{Aonq}), or they involve 
relativistic $P$-states.

Using Eqs.(\ref{g1N}), (\ref{eq:g1A}) and (\ref{Aonq}) we can decompose
$g_1^D$ into a convolution component plus an off-shell correction:
\begin{eqnarray}
\label{comps}
g_1^D(x)
= \int_x^1\ {dy \over y } \Delta f(y)\ g_1^N\left({x\over y}\right)\
 +\ \delta^{(OFF)} g_1^D(x). 
\end{eqnarray}
Here we can identify 
\begin{eqnarray}
\label{Dfy}
\Delta f(y)
= \int d^4p\
  \Delta S(p)\
  \delta \left( y - \frac{p^+}{M_D} \right)
\end{eqnarray}
with the difference of probabilities
to find a nucleon in the deuteron with light-cone momentum fraction
$y$ and spin parallel and antiparallel to that of the deuteron.
For a deuteron with polarization $m=+1$,
$\Delta S(p)$ corresponds to the spectral function:
\begin{eqnarray}
\label{DSp}
\Delta S(p)
= \Psi_{+1}^{\dagger}(p)\
  \left( \widehat{\cal S}_0\ +\ \widehat{\cal S}_z \right)
  \Psi_{+1}(p)\ \
  \delta\left( p_0 - M_D + E_p \right),
\end{eqnarray}
where $\Psi_m(p)$ is the usual (normalized) non-relativistic deuteron
wave function (see e.g. Ref.\cite{DNR}). 
$\widehat{\cal S}_0$ and $\widehat{\cal S}_z$ are
the zero and $z$ components  of the nucleon spin operator, defined
as \cite{KMPW}:
\begin{mathletters}
\label{Shat}
\begin{eqnarray}
\widehat{\cal S}_0
&=& \frac{{\bf S \cdot p}}{M} 
= \frac{1}{2M} \left(\mbox{\boldmath$\sigma$}^p +
                    \mbox{\boldmath$\sigma$}^n\right)
\cdot {\bf p},  \\
\widehat{\cal S}_z
&=& \frac{1}{2} \left( \sigma^p_z + \sigma^n_z\right),
\end{eqnarray}
\end{mathletters}%
with $\mbox{\boldmath$\sigma$}^{p,n}$ the SU$(2)$ Pauli spin 
matrices acting on 
the proton and neutron spin wave function, respectively.
In terms of the deuteron wave function 
$\Psi_m(p)$, one has:  
\begin{eqnarray}
\label{PsiNR}
q \cdot {\cal A}_{ON} = {\cal T\cdot B}_{ON}  
= 8 \pi^3 P\cdot q M \ {\cal N}\ \
  \Psi_{+1}^{\dagger}(p)\
  \left( \widehat{\cal S}_0 + \widehat{\cal S}_z \right)\
  \Psi_{+1}(p).
\end{eqnarray}
The factor 
${\cal N} = \int d|{\bf p}|\ {\bf p}^2\ (u^2 + w^2)$
ensures that the normalization agrees with that of the re\-la\-tivistic
calculation.
The function $\Delta f(y)$ then satisfies
$ \int_0^1 dy\ \Delta f(y)
= 1 - 3/2\ \omega_D $,
where
$\omega_D = \int d|{\bf p}|\ {\bf p}^2\ w^2\ /\ {\cal N}$
is the non-relativistic $D$-state probability.
In the $NN$ potential model of
Ref.\cite{BG} with a pseudovector $\pi NN$ interaction, 
the $D$-state probability is $\omega_D = 4.7\%$.
The convolution term in Eq.(\ref{comps}) completely agrees with 
the non-relativistic limit up to order $(v/c)$ if contributions from 
relativistic $P$-states are neglected, and the off-shell 
structure function $\tilde g_1^N(x/y,p^2)$ from Eq.(\ref{g1N}) 
is replaced by the on-shell one. 
The relativistic, convolution breaking, ``off-shell'' 
contribution $\delta^{(OFF)} g_1^D$ in Eq.(\ref{comps}) 
is given explicitly in Appendix \ref{OFF}.

We should also make a note about comparing calculations which use
relativistic and non-relativistic wave functions.
While convolution itself is valid to order $(v/c)^2$ \cite{KMPW},
the renormalization of the relativistic wave function itself 
introduces corrections of order $(v/c)^2$, since the $P$-state 
wave functions $v_{s,t}$ are of order $v/c$ compared with the 
$S$- and $D$-state functions.
Therefore the correct non-relativistic limit can be obtained
directly from the relativistic calculation only to order $v/c$
\cite{SK}.

\section{Numerical Results}

Using the results of the preceeding sections, we present here the 
numerical results for the 
polarized nucleon and deuteron structure functions.
In our relativistic treatment, the modeling of the virtual
photon--off-shell nucleon interaction is most naturally done
in terms of relativistic quark--nucleon vertex  functions.

\subsection{Relativistic Quark---Nucleon Vertex Functions}

While the scaling behavior of $G_{(i)}$ can be derived from the 
parton model, their complete evaluation requires model-dependent 
input for the non-perturbative parton--nucleon physics, which in 
our case is parametrized by the functions $f_{1\cdots 10}$ 
and $g_{1\cdots 9}$.
Because the nucleon recoil state that remains a spectator to the 
hard collision is on-mass-shell, the functions 
$f_{1\cdots 10}, g_{1\cdots 9}$ can be
simply written in terms of relativistic quark--nucleon vertex 
functions, ${\cal V}$ \cite{MST,MSTD,MPT,MM,MW,KMWW}.
Expressed through $\cal V$ and the propagator of the 
spectator (``diquark'') system, $S_D(p-k)$, one obtains for 
the matrix ${\cal H}(k,p)$ from Eq.(\ref{Ghatlead}): 
\begin{eqnarray} \label{H}
{\cal H}(k,p)
&=& Im \left[ {\cal V}(k,p) S_D(p-k)
              \overline{{\cal V}}(k,p)\right ],
\end{eqnarray}
where $ \overline{{\cal V}} = \gamma_0 {\cal V}^{\dagger} \gamma_0$. 
Because both the quark and nucleon have spin 1/2, the spectator 
system can be in either a spin $0$ or spin $1$ state.
Therefore the only vertices that need to be considered (for quarks
in the ground state) are those which transform as scalars or 
pseudovectors under Lorentz transformations.
We approximate the ``diquark'' propagators by the 
form: 
$S_D(p-k) = \left( (p-k)^2 - m_0^2 \right)^{-1}$ for $S=0$, 
and 
$S_D^{\alpha\beta}(p~-~k) 
= \left( -g^{\alpha\beta} + (p-k)^{\alpha}(p-k)^{\beta}/m_1^2 \right)
/ \left( (p-k)^2 - m_1^2 \right)$ 
for $S=1$, where $m_0$ and $m_1$ are the masses associated with  
the scalar and pseudovector spectator states.

Following earlier work \cite{MPT}, we will use the  
ansatz:
\begin{mathletters}
\begin{eqnarray}
\label{Vertices}
&{\cal V}&_0(k,p) = I \phi_0^{(a)}(k,p)
                  + \beta \overlay{\slash}k \phi_0^{(b)}(k,p),  \\
&{\cal V}&_1^{\alpha}(k,p) = \gamma^{\alpha} \gamma_5 \phi_1(k,p),
\end{eqnarray}
\end{mathletters}%
for the $S=0$ and $S=1$ vertices, respectively.
The parameter $\beta$ determines the relative contributions from 
the two scalar vertices.
Although a more general approach is possible, in which one could
include all possible Lorentz and Dirac structures, in practice
since the vertices will be constrained phenomenologically, the 
above set will suffice.
Without loss of generality, to simplify the numerical analysis 
we will also work in the massless quark limit, $m \rightarrow 0$.

Inserting the above vertex functions into Eq.(\ref{Ghatlead})
and comparing with Eq.(\ref{Ghatex}), we can determine
the functions $f_{1\cdots 10}$, $g_{1\cdots 9}$ appearing 
in Eqs.(\ref{Hdef}).
For the scalar vertices we find:
\begin{mathletters}
\begin{eqnarray}
g_4 &=& -2 \beta^2 \left( \phi_0^{(b)} \right)^2,        \nonumber\\
g_7 &=& -\left( \phi_0^{(a)} \right)^2\
        - k^2 \beta^2 \left( \phi_0^{(b)} \right)^2,    \\
g_9 &=& 2 \beta \phi_0^{(a)} \phi_0^{(b)},              \nonumber
\end{eqnarray}
while for the pseudovector vertex:
\begin{eqnarray}
g_1 &=& - {2 \over m_1^2} \left( \phi_1 \right)^2\
     =\ -g_2\ =\ -g_3\ =\ g_4\
     =\ -{2 \over m_1^2} g_7,
\end{eqnarray}
\end{mathletters}%
with all other functions being zero.

The momentum dependence in the vertices is parametrized by the 
multipole form:\
$ \Phi_S(p,k)
= N_S(p^2)\ \cdot\ k^2 / (k^2 - \Lambda_S^2)^{n_S}$
($p^2 = M^2$ for the free nucleon).
The cut-off parameters $\Lambda_S$ and exponents $n_S$
are fixed by fitting the unpolarized up ($u_V$) and down ($d_V$) 
valence quark distributions, as discussed in Refs.\cite{MST,MPT}.
The normalization constants $N_S$ are determined through baryon number 
conservation.
A best fit to the experimental $u_V + d_V$ and 
$d_V/u_V$ data \cite{PARAM} at 
$Q^2 = 10$ GeV$^2$ (when evolved from the renormalization scale
$\mu^2 \simeq$ (0.32 GeV)$^2$
using leading order evolution\footnote{
While a next-to-leading order analysis is important
for a precise determination of the $Q^2$-dependence of $g_1$
and the Bjorken sum rule \cite{Q2},
the present treatment is perfectly adequate for the purpose
of evaluating the relative sizes of the nuclear corrections.}),
is shown in Fig.2.
The cut-offs used for the scalar vertices are 
$\Lambda_0^{(a,b)} = (1.0, 1.1)$ GeV, and the exponents
$n_0^{(a,b)} = (2.0,2.8)$, 
with the mixing parameter $\beta = 2.73$.
The parameters for the pseudovector vertex are $\Lambda_1 = 1.8$ GeV 
and $n_1 = 3.2$.
The mass parameters associated with the intermediate spectator
states are taken to be $m_{0(1)} = (p-k)^2 = 0.9 (1.6)$ GeV.

With the same parameters, the polarized valence distributions are then 
calculated according to the relations\footnote{
Note that the formal results do not rely on the use of SU(4) 
spin-flavor symmetry --- these relations merely provide a 
convenient way to parametrize the polarized quark distributions
\cite{MM}.}:
\begin{eqnarray}
\Delta u_V (x)
&=& \frac{3}{2}\Delta q_0(x)\
           -\ \frac{1}{6} \Delta q_1(x), \\
\Delta d_V (x)
&=& - \frac{1}{3}\Delta q_1(x),
\end{eqnarray}
where $\Delta q_{0}$ and $\Delta q_{1}$ are the polarized quark 
distributions for scalar and pseudovector spectator states,   
respectively. 
The first moments of the polarized valence distributions 
in the proton then turn out to be:\   
$\int_0^1 dx\ \Delta u_V(x) = 0.99$ and
$\int_0^1 dx\ \Delta d_V(x) = -0.27$,
which saturates the Bjorken sum rule:
$ \int_0^1 dx\ (\Delta u_V(x) - \Delta d_V(x))
= g_A $.
The total momentum carried by valence quarks at the scale $\mu^2$ 
is around 85\%, leaving about 15\% to be carried by the sea.

The $x$-dependence of the (valence part of the) polarized 
proton structure function
$ x g_1^p(x)
= x (4 \Delta u_V(x) + \Delta d_V(x)) / 18$\
is shown in Fig.3.
In the valence quark dominated region ($x > 0.3$) the
result agrees rather well with the SLAC, EMC and SMC
proton data \cite{SLAC,EMC,SMCP,E143P}.
A negatively polarized sea component at $x < 0.3$ would
bring the curve even closer to the data points.

Having fixed the nucleon inputs, we next estimate the
size of the relativistic corrections to the deuteron structure
function.

\subsection{Polarized Deuteron Structure Function}

The total valence part of the structure function of the deuteron,
calculated from Eq.(\ref{comps}), is shown in Fig.4.
The agreement with the SMC \cite{SMCD} and SLAC E143 \cite{E143D} 
data in the valence region is also quite good.
Note that care must be taken regarding the normalization of the
quark-nucleon vertex functions when the nucleon is no longer 
on-shell.
In this simple model, the modifications of the vertex functions
are made via the explicit $p^2$ dependence of the normalization 
constants $N_S(p^2)$. 
In practice, because the structure function is not very sensitive 
to the $p^2$ dependence in the quark--nucleon vertex functions,
$N_S(p^2)$ can be taken to be constant.
The numerical values of these normalization constants are fixed 
through valence quark number conservation in the deuteron's
spin-averaged distributions. 
They turn out to be 0.8\% and 1.7\% smaller for the scalar and 
pseudovector vertices, respectively, than for the free nucleon. 
The insensitivity of the deuteron structure function to the
$p^2$ dependence of the quark--nucleon vertex should limit the
uncertainty introduced through the specific choice of vertex
functions in Eqs.(\ref{Vertices}).
Any model dependence associated with alternative vertex structures
would be compensated to some extent by the necessary adjustments 
to their momentum dependence --- the $k^2$ dependence is constrained 
by refitting the nucleon data, and the $p^2$ dependence by readjusting
the normalization constants $N_S$.
Nevertheless, it would be interesting to explore the residual model
dependence numerically.

The resulting ratio, $g_1^D/g_1^N$, is displayed in Fig.5.
For large $x$ it exhibits the same characteristic shape
as for the (unpolarized) nuclear EMC effect, namely a dip
of $\sim$ 7-8\% at $x \approx 0.6$ and a steep rise due to
Fermi motion for $x > 0.6$.
For small $x$ it stays below unity, where it can be reasonably
well approximated by a constant depolarization factor,
$1 - 3/2\ \omega_D$, as is typically done in data analyses
\cite{SMCD,E143D}.
Also shown in Fig.5 is the ratio of the convolution ansatz
(Eqs.(\ref{comps}) -- (\ref{DSp})) to the full calculation
(dashed curve).
As one can see, this ansatz works remarkably well for most $x$,
the only sizable deviations occuring for $x > 0.8$,
which is outside the range covered by previous
experiments. 
Nevertheless, future experiments, both inclusive and  
semi-inclusive, will be able to access the very large-$x$
region, in which case the issue of nuclear --- and in particular
relativistic --- effects will need to be seriously addressed.

\section{Conclusion}

We have discussed polarized nuclear deep-inelastic scattering 
within a covariant framework. 
In this context we analyzed the structure  of the forward 
scattering amplitude of a virtual photon from a bound,
off-mass-shell nucleon, focussing especially on its symmetry 
properties. 
Within the impulse approximation, 
we derived  the most general, relativistic expression for the 
nuclear structure function $g_1^A$.  
Our results clearly demonstrate that, in general,
nuclear and nucleon pieces do  
not factorize in a relativistic treatment of nuclear structure 
functions. 
The conventional convolution model can only be recovered 
by taking the non-relativistic limit for the nucleon--nucleus 
amplitude and assuming the on-shell limit for  the off-shell 
nucleon structure function.

We showed numerical results for the deuteron structure function
$g_1^D$, where the off-shell nucleon tensor was calculated within 
a relativistic spectator model. 
At moderate $x$ ($ <0.7$) we found that nuclear effects in the deuteron 
are dominated by the $D$-state probability of the deuteron. 
Binding and Fermi-motion become significant only at large $x$. 
Also off-shell corrections turned out to be small at moderate $x$,  
but increase considerably in the region $x>0.8$.

With respect to the present experimental situation the 
main uncertainty in the extraction of the neutron structure function 
$g_1^n$ still comes from the deuteron $D$-state probability. For
practical purposes, the off-shell modification of the bound
nucleon structure function has to be taken into account 
when high precision data at large $x$ become available.

\acknowledgements

We would like to thank S.Kulagin and W.Weise
for helpful comments and discussions.
This work was supported in part by
the Australian Research Council,  
BMBF and the U.S. Department of Energy
grant \# DE-FG02-93ER-40762.

\appendix
\section{Truncated structure functions}
\label{Gi}

Truncated structure functions $G_{(i)}$ from Eq.(\ref{Ghatex}) 
in terms of the functions $f_{1\cdots 10}$ and $g_{1\cdots 9}$
Eq.(\ref{Hdef}): 

\begin{eqnarray}
G_{(p)}\left( {x\over y}, p^2 \right)
&=& \int d\widetilde{k}
\left\{ \left( m^2 + k^2 - 2 p\cdot k {x \over y}
        \right)
        \left( g_1 + {x\over y} g_2 \right)
     -\ {x \over y} (k^2 - m^2)
        \left( g_3 + {x\over y} g_4 \right)
\right.                                         \nonumber\\
& &
\left.
     +\ 2 \left( {x \over y} \right)^2 g_7\
     +\ 2 m \left( k^2 - p\cdot k {x \over y} \right)
        \left( f_7 + {x\over y} f_8 \right)
     +\ 2 m {x \over y} \left( f_9 + {x \over y} f_{10} \right)
\right\},                                       \\                     
G_{(q)}\left( {x\over y}, p^2 \right)
&=& \int { d\widetilde{k} \over p\cdot q }
\left\{ \left( (m^2 + 2 k^2) p\cdot k
             - \left( (m^2 + k^2) p^2 + 4 (p\cdot k)^2
               \right) {x \over y}
             + 3\ p^2\ p\cdot k\ \left( {x \over y} \right)^2
        \right) g_2\
\right.                                         \nonumber\\
& &
     +\ \left( k^2 - 4\ p\cdot k {x \over y}
                   + 3 p^2 \left( {x \over y} \right)^2
        \right)
        \left( {1\over 2} (k^2 - m^2) g_4 - m f_{10}
        \right)                                 \nonumber\\
& &
     +\ m \left( 3 k^2 p\cdot k
               - 2 \left( p^2 k^2 + 2 (p\cdot k)^2 \right) {x \over y}
               + 3 p^2\ p\cdot k \left( {x \over y} \right)^2
          \right) f_8\                          \nonumber\\
& &
\left.
     +\ 2 m \left( p\cdot k - p^2 {x \over y}
            \right) f_9\
     -\ \left( k^2 - 4 p\cdot k {x \over y}
                   + 3 p^2 \left( {x \over y} \right)^2
        \right) g_7\
\right\}                                                        \\
G_{(\gamma)}\left( {x\over y}, p^2 \right)
&=& \int d\widetilde{k}
\left\{ \left( k^2 - 2 p\cdot k {x \over y}
             + p^2 \left( {x \over y} \right)^2
        \right)
        \left( - p\cdot k\ g_2\
               - {1\over 2} (k^2-m^2) g_4
               + g_7
        \right)\   
\right.                                     \nonumber \\
& &
\left.    
  -\ (m^2 + k^2) g_7\
     +\ m \left( k^2 - 2 p\cdot k {x \over y}
               + p^2 \left( {x \over y} \right)^2
          \right)
          \left( - p\cdot k\ f_8\ + f_{10} \right)
\right.                                     \nonumber \\
& &
\left.  
     -\ 2 m \left( p\cdot k\ f_9 + k^2 f_{10} \right)
\right \}                                                        \\
G_{(\sigma p)}\left( {x\over y}, p^2 \right)
&=& \int d\widetilde{k}
\left\{ \left( k^2 - 2 p\cdot k {x \over y}
             + p^2 \left( {x \over y} \right)^2
        \right)
        \left( - p\cdot k\ g_5
               - {1\over 2} (k^2-m^2) g_6
               - g_8
        \right)
\right.                                         \nonumber\\
& &
        +\ (k^2 + m^2) \left( g_8 + {x \over y} g_9 \right)\
        -\ 2 m \left( p\cdot k - p^2 {x \over y}
               \right) f_2\
        -\ m \left( k^2 - p^2 \left( {x \over y} \right)^2
             \right) f_3\                       \nonumber\\
& &
\left.
        +\ 2 m {x \over y} f_6\
\right\}                                                \\
G_{(\sigma q)}\left( {x\over y}, p^2 \right)
&=& \int { d\widetilde{k} \over p\cdot q }
\left\{ \left( p\cdot k - p^2 {x \over y} \right)
        \left( (k^2 + m^2) g_9\
              + 2 m \left( p^2 f_2 + p\cdot k f_3 + f_6 \right)
        \right)
\right\}                                                \\
G_{(\sigma pq)}\left( {x\over y}, p^2 \right)
&=& \int { d\widetilde{k} \over p\cdot q }
\left\{ - \left( p\cdot k\ (m^2 + 2 k^2)
               - \left( (k^2 + m^2) p^2 + 4 (p\cdot k)^2
                 \right) {x \over y}
               + 3 p^2 p\cdot k \left( {x \over y} \right)^2
          \right) g_5\
\right.                                         \nonumber\\
& &
        +\ \left( k^2 - 4 p\cdot k {x \over y}
                      + 3 p^2 \left( {x \over y} \right)^2
           \right)
           \left( - {1\over 2} (k^2-m^2) g_6
                  - g_8
                  + m f_3
           \right)                              \nonumber\\
& &
\left.
        -\ 2 m \left( p\cdot k - p^2 {x \over y}
               \right) f_2\
\right\}
\end{eqnarray}
where $p \cdot k = (p^2 + k^2 - m_S^2)/2$ and 
$x/y = k \cdot q/p \cdot q = Q^2/2 p\cdot q$.

\section{Relativistic corrections to nuclear structure functions}
\label{OFF}

As outlined in Sec. IV the pseudovector and tensor components of the 
nucleon--deuteron scattering amplitude, ${\cal A}^{\alpha}$ and 
${\cal B}^{\alpha \beta}$ in  Eq.(\ref{ApPS}), can be separated into
an on-shell part, which is proportional to the non-relativistic 
deuteron wave function, and an off-shell component:  
$q\cdot {\cal A}\equiv q\cdot {\cal A}_{ON} + q\cdot {\cal A}_{OFF}$, 
and 
${\cal T\cdot B} \equiv {\cal T\cdot B}_{ON} + {\cal T\cdot B}_{OFF}$.
The on-shell component $q\cdot {\cal A}_{ON} = {\cal T \cdot B}_{ON}$ 
is given in Eq.(\ref{Aonq}). The off-shell contributions,
which are of higher order in $v/c$ or involve relativistic 
$P$-states, are 
\begin{eqnarray}
q\cdot {\cal A}_{OFF}
&=& 2 \pi^2 P\cdot q M
    \left[
     \cos^2\theta \left( { E_p \over M } - 1 \right)
    \left( u - { w \over \sqrt{2} } \right)^2
    \right.                                                     \nonumber\\
& & -\ {3 \over 2}
       \left( {p_z \over M}
            - { E_p \over M } \cos^2\theta
       \right) v_t^2\
 +\ {3 \over \sqrt{2}} \left( 1 - \cos^2\theta \right) v_s v_t  \nonumber\\
& & +\ \sqrt{6} \left( \cos\theta
                     - {|{\bf p}| \over 2 M}
                       \left( 1 - \cos^2\theta \right)
                \right) u v_t\
    -\ \sqrt{3} \left( \cos\theta
                     - {|{\bf p}| \over M}
                       \left( 1 - \cos^2\theta \right)
                \right) w v_t\                                 \nonumber\\
& & \left.
    +\ {\sqrt{3} |{\bf p}| \over M} \left( 1 - \cos^2\theta \right)
       \left( u - {w\over\sqrt{2}} \right) v_s
    \right],                                             \label{Aoffq}
\end{eqnarray}
and
\begin{eqnarray} \label{BoffT}
{T\cdot \cal B}_{OFF}
&=& 2\pi^2 P\cdot q M^2
\left[
\left( {\varepsilon\over M}
   + {{\bf p}^2 \over M (M + E_p) }
     \left( - \cos^2\theta
          + {\varepsilon - 2 E_p \over M} (1 - \cos^2\theta)
     \right)
\right) u^2
\right.                                         \nonumber\\
& & +
\left(  {E_p - M \over M} - {{\bf p}^2\over M^2} 
          (1-\cos^2\theta) 
       - {\varepsilon \over 2 M}  
          {E_p - \cos^2\theta (E_p + 2 M) \over M}
\right) \sqrt{2} u w                                     \nonumber\\
& & +
\left(  {M-E_p \over M} \left(2 - {3\over2} \cos^2\theta\right)
      + 2 {{\bf p}^2\over M^2} (1-\cos^2\theta)
     -  {\varepsilon\over M} 
        {E_p - \cos^2\theta (E_p +  M) \over M}
\right) w^2                                     \nonumber\\
& & +
{3\over 2}
\left( -{p_z\over M} + {E_p - \varepsilon - 2 M \over M} \cos^2\theta
\right) v_t^2                                   \nonumber\\
& & +
{3\over \sqrt{2}}
\left( 1 -  {E_p \over M} \left( 2 +  {\varepsilon \over M}\right)
\right) (1-\cos^2\theta) v_s v_t                \nonumber\\
& & +
\sqrt{3\over 2} {|{\bf p}|\over M}
\left( - 2 {p_z\over M} + 2 {E_p - \varepsilon \over M} \cos^2\theta
       - 1 - 3 \cos^2\theta
\right) u v_t                                   \nonumber\\
& & +
\sqrt{3} {|{\bf p}|\over M}
\left( {p_z\over M} - {E_p - \varepsilon \over M} \cos^2\theta
       - 1 + 3 \cos^2\theta
\right) w v_t                                   \nonumber\\
& & \left. +
\sqrt{3} {|{\bf p}|\over M} (1-\cos^2\theta)
\left( u - {w \over \sqrt{2}} \right) v_s
\right],
\end{eqnarray}
where 
$p_z = |{\bf p}| \cos\theta$,
$E_p = \sqrt{ M^2 + {\bf p}^2 }$ and
$\cos\theta = (y M_D - p_0) / |{\bf p}|$.
Also the convolution breaking contribution $p\cdot{\cal A}$ 
in Eq.(\ref{eq:g1A}) 
represents a relativistic correction in comparison with the 
``on-shell'' amplitude:
\begin{eqnarray} \label{Ap}
{\cal A}\cdot p 
&=& 2 \pi^2 M_D M
       \left[ \left( M_D - 2 E_p \right)
              { p_z \over M }
              \left( u - { w \over \sqrt{2} } \right)^2
           -\ {3 M_D p_z \over 2 M } v_t^2\
       \right.                                    \nonumber\\
& & \hspace*{0cm} \left.
           +\ \sqrt{6} \left( M_D - E_p \right) \cos\theta
              \left( u - {w \over \sqrt{2}} \right) v_t
       \right].                      
\end{eqnarray}

Taking into account these relativistic contributions to the 
nucleon--deuteron amplitude, one obtains the relativistic 
correction to the convolution component of $g_1^D$ in 
Eq.(\ref{comps}): 
\begin{eqnarray} 
\delta^{(OFF)} g_1^D(x) 
&=& \frac{P\cdot q}{4 \pi^2 M_D S\cdot q}
\int dy\,dp^2\, 
\Bigg[ \frac{q\cdot{\cal A}_{ON}}{2 p\cdot q} \
\left(\tilde g_1^N\left({x}/{y},p^2\right) 
- g_1^N(x/y)\frac{E_p}{{\cal N} M}\right)\nonumber \\
&+&  
q\cdot {\cal A}_{OFF} \left( p\cdot q \,G_{(q)} + G_{(\gamma)} \right) 
+ {{\cal T\cdot B}_{OFF}} M \left( G_{(\sigma p)} - 
p \cdot q \,G_{(\sigma pq)} \right) 
 \nonumber \\
&+& p\cdot {\cal A} \,p\cdot q \,G_{(p)} \Bigg].
\end{eqnarray}
%


\begin{figure}
\centering{\ \psfig{figure=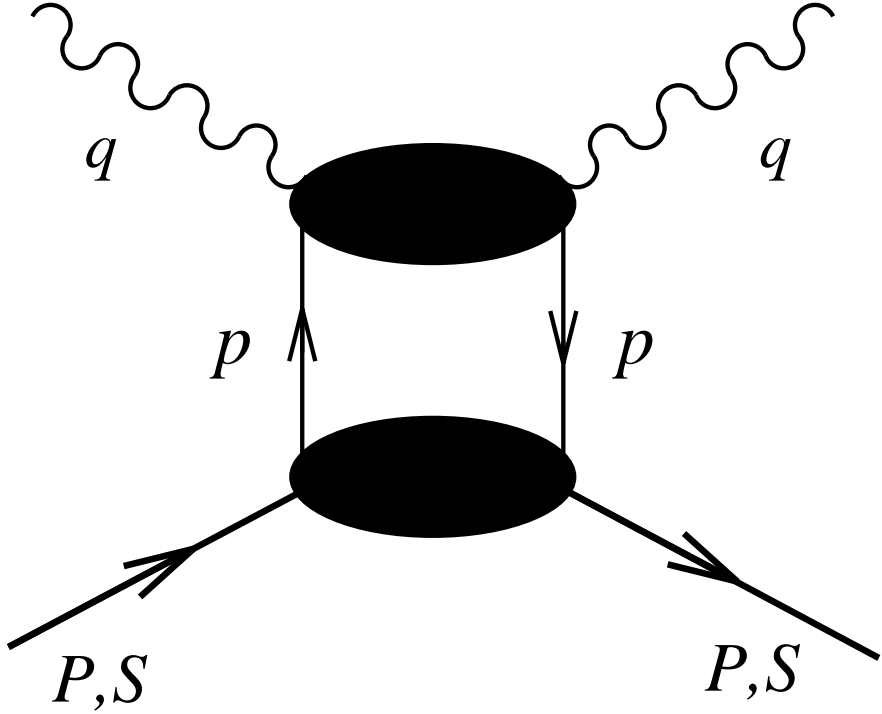,height=13cm}}
\vspace*{1cm}
\caption{DIS from a polarized nucleus  in the impulse approximation.
        The nucleus, virtual nucleon and photon momenta are denoted
        by $P$, $p$ and $q$, respectively, and $S$ stands for 
        the nuclear spin vector. The upper blob represents the 
        truncated antisymmetric nucleon tensor $\widehat G_{\mu\nu}$,  
        while the lower one corresponds to the polarized nucleon--nucleus 
        amplitude $\widehat A$.}
\end{figure}

\begin{figure}
\centering{\ \psfig{figure=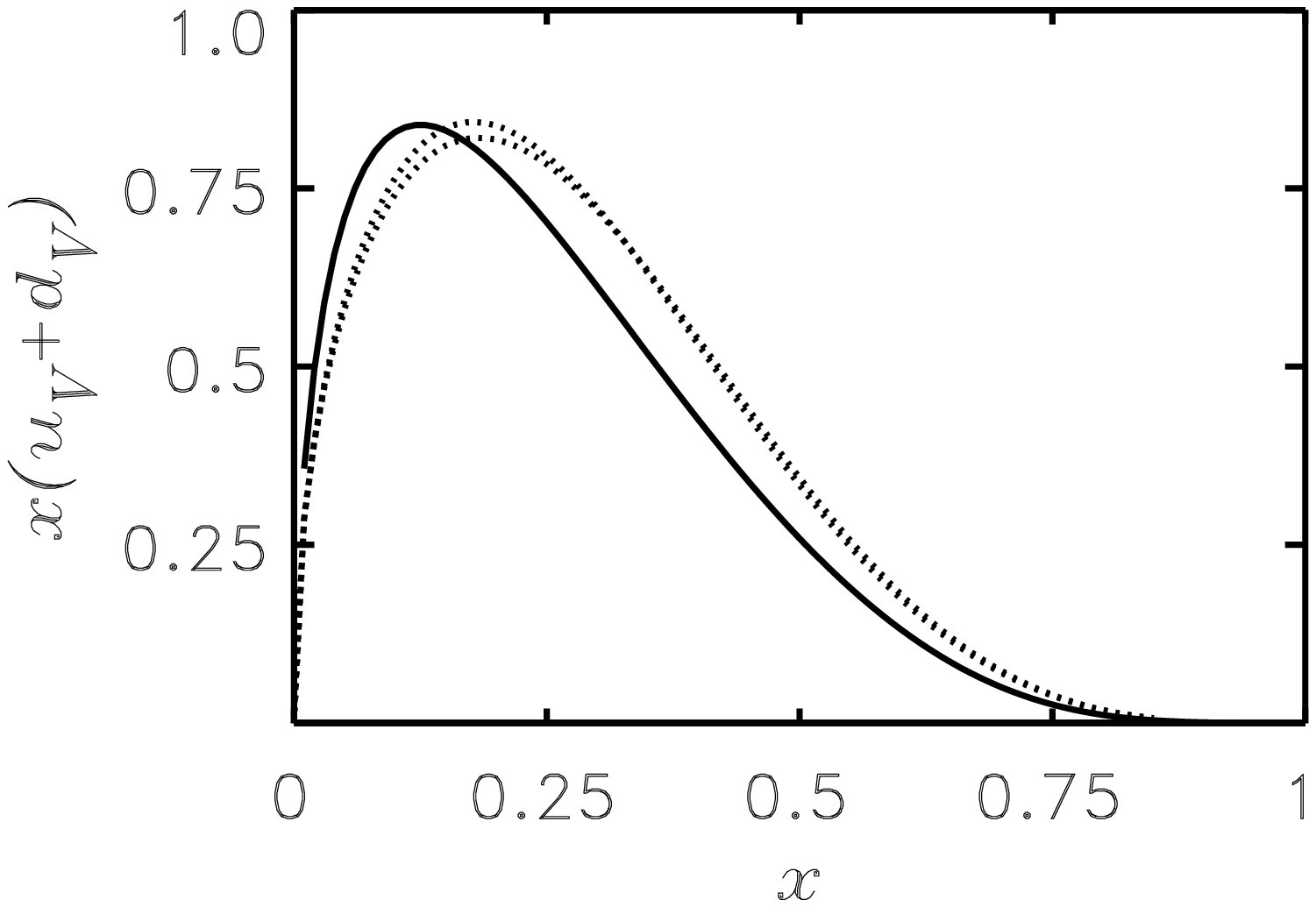,height=10cm}}
\centering{\ \psfig{figure=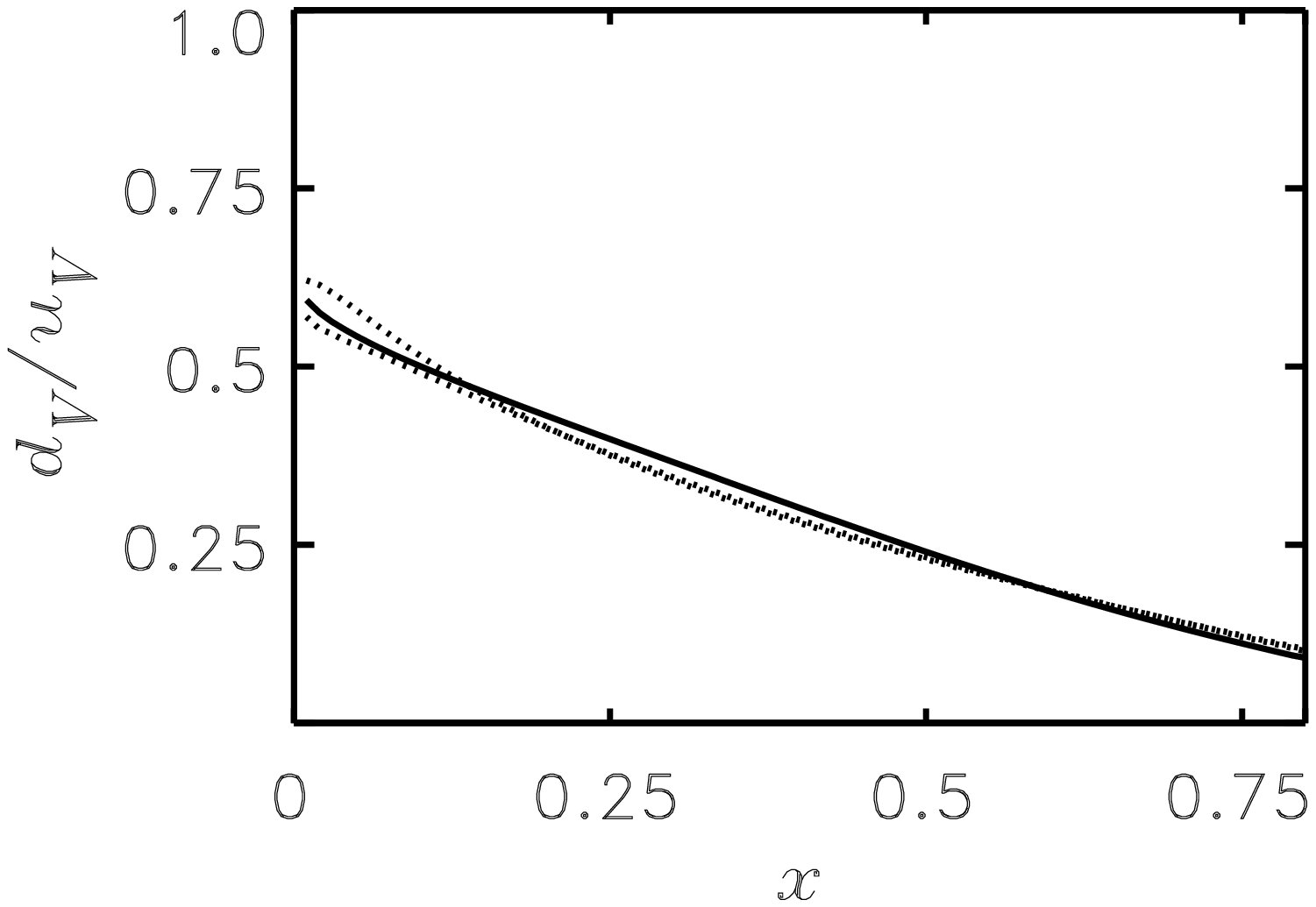,height=10cm}}
\caption{Unpolarized valence quark distributions $x(u_V+d_V)$ and 
        $d_V/u_V$.  The solid line represents distributions 
        evolved from scale $\mu^2$ to $Q^2=10$ GeV$^2$.
        Dotted curves are parametrizations from 
        Ref.\protect\cite{PARAM}.}   
\end{figure}

\begin{figure}
\centering{\ \psfig{figure=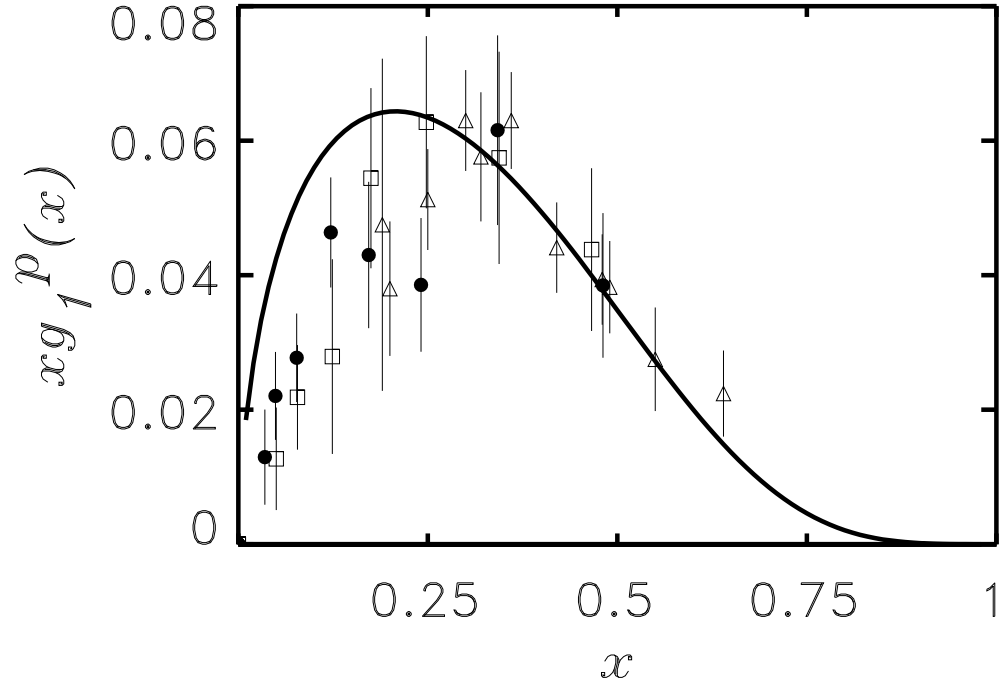,height=11cm}}
\caption{Valence component of the proton 
        $g_1$ structure function at $Q^2=10$ GeV$^2$.
        The data for the full structure function $g_{1}^p$ are from 
        Refs.\protect\cite{SLAC,EMC,SMCP,E143P}.}  
\end{figure}

\begin{figure}
\centering{\ \psfig{figure=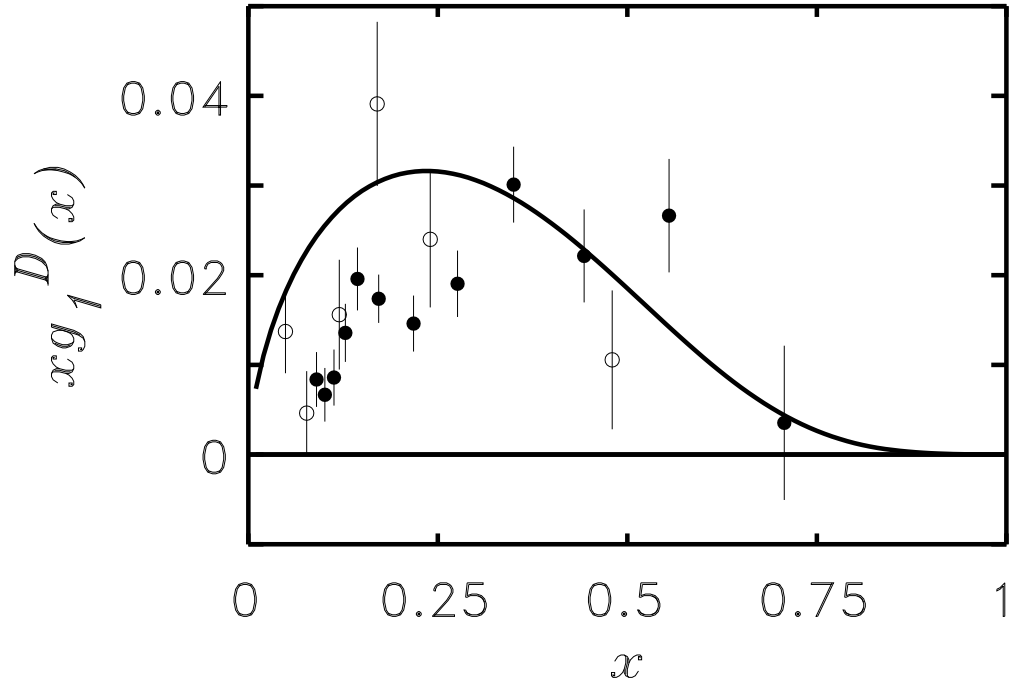,height=11cm}}
\caption{Valence part of the deuteron structure function $g_{1}^D$, 
        compared with the data for the full $g_1^D$ structure 
        function from the SMC  and SLAC E143 \protect\cite{SMCD,E143D}.}
\end{figure}

\begin{figure}
\centering{\ \psfig{figure=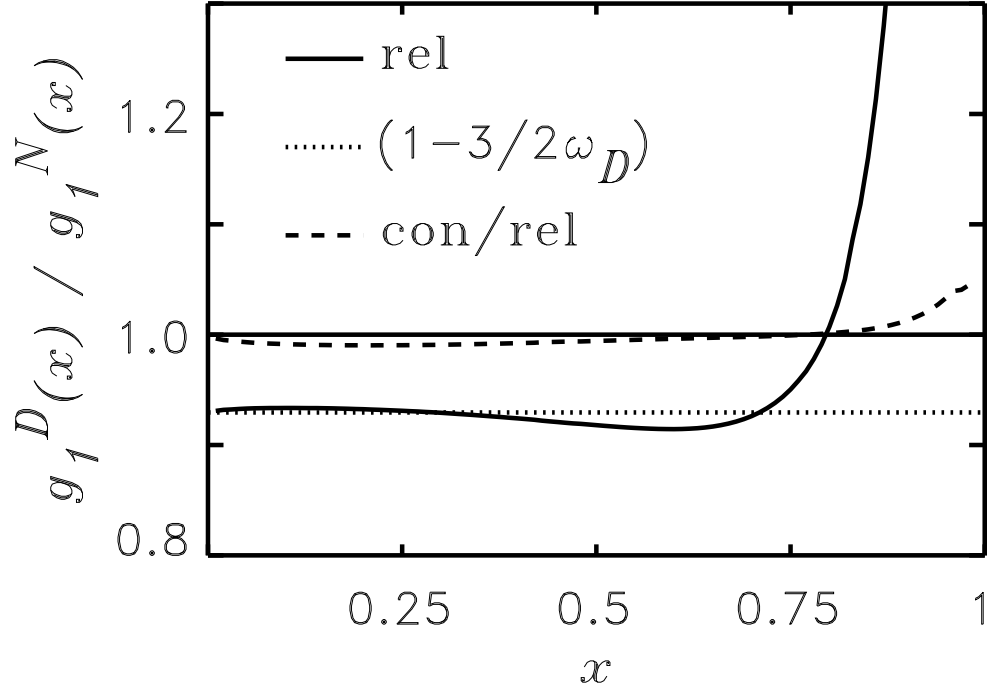,height=11cm}}
\caption{Ratio of deuteron and nucleon structure functions in the
        full model (solid), and with a constant depolarization factor
        $1 - 3/2\ \omega_D$ (dotted), with $\omega_D = 4.7\%$ 
        \protect\cite{BG}.  
        Dashed curve is ratio of $g_1^D$ calculated via convolution
        to $g_1^D$ calculated in the relativistic model.}  
\end{figure}

\end{document}